\documentclass[]{aa} % for a printer version

\newcommand{\msun}{$\mathrm{M_{\odot}}$}
\newcommand{\teff}{T$_{\rm eff}$}

\usepackage{color}
\usepackage{soul}

\usepackage{txfonts}
\usepackage{graphicx}
\usepackage[authoryear]{natbib}
\usepackage{epsfig}

\usepackage{natbib}

\bibpunct{(}{)}{;}{a}{}{,} % to follow the A&A style

\begin{document}

\title{Age and mass of solar twins constrained by lithium abundance
\thanks{The  models  are only available in electronic form at 
the CDS via anonymous ftp to  cdsarc.u-strasbg.fr (130.79.128.5) or via 
http://cdsweb.u-strasbg.fr/cgi-bin/qcat?J/A+A/... or via http://andromeda.dfte.ufrn.br}
%\fnmsep
}
 \subtitle{}

   \author{J.D. do Nascimento Jr\inst{1},
           M. Castro\inst{1},
           J. Mel\'endez\inst{2},
           M. Bazot\inst{2},
           S. Th\'eado\inst{3},
           G. F. Porto de Mello\inst{4},
           J.R. De Medeiros\inst{1}
}

   \offprints{J.D. do Nascimento, email:dias@dfte.ufrn.br}

\institute{Departamento de F\'isica Te\'orica e Experimental,
Universidade Federal do Rio Grande do Norte, CEP: 59072-970 Natal,
RN, Brazil \and Centro de Astrof\'isica da Universidade do Porto,
Rua das Estrelas, 4150-762 Porto, Portugal \and Laboratoire
d'Astrophysique de Toulouse et Tarbes - UMR 5572 - Universit\'e
Paul Sabatier - CNRS, 14 Av. E. Belin, 31400
Toulouse, France \and Universidade Federal do Rio de Janeiro,
Observat\'orio do Valongo, Ladeira do Pedro Ant\^onio, 43, Rio de
Janeiro, CEP: 20080-090, Brazil}

\date{Received Date: Accepted Date}

\abstract
% Context
{}
% Aims:
{We analyze the non-standard mixing history of the solar twins HIP
55459, HIP 79672, HIP 56948, HIP 73815, and HIP 100963,  to
determine as precisely as possible their mass and age.}
% Methods:
{We  computed a grid of evolutionary models with non-standard
mixing at several metallicities with the Toulou\-se-Geneva code for
a range of stellar masses assuming an error bar of $\pm$50K in
\teff. We choose the evolutionary model that reproduces accurately the
observed low lithium abundances observed in the solar twins.}
% Results:
{Our best-fit model for each solar twin provides a mass and age
solution constrained by their Li content and \teff~determination. 
HIP 56948 is the most likely solar-twin candidate  at the present
time and our analysis infers a mass of $0.994 \pm 0.004$ \msun \
and an age of $4.71 \pm 1.39$ Gyr.}
% Conclusions:
{Non-standard mixing is required to explain the low Li abundances
observed in solar twins. Li depletion due to additional mixing
in solar twins is strongly mass dependent.  An accurate
lithium abundance measurement and  non-standard models
provide  more precise information about the age and mass
more robustly  than determined  by classical methods alone.} 
\titlerunning{Age and mass of solar twins constrained by lithium abundance}
\authorrunning{do Nascimento et al.}

\maketitle

\section{Introduction}
\label{sec:Intro}

Lithium is easily destroyed by nuclear burning in stellar
interiors at temperatures above $2.4 \times 10^6$ K and its
surface abundance in main-sequence stars indicates the depth of
mixing below their photospheres. As in the Sun, the amount of Li
depletion in solar twins is sensitive to microscopic diffusion,
and some extra-mixing process is required to explain the low
observed Li abundances, indicating that they also share with the
Sun a similar mixing history.  Standard
solar models predict that the Sun has a convective envelope whose
mass is only 2\% of the total mass and where the base temperatures
are generally too low to destroy lithium. The solar Li
problem is the long-standing conflict between the observed
photospheric Li depletion of the Sun by 2.21 dex (Grevesse \&
Sauval 1998) and the low theoretically predicted depletion
of stellar evolution models based only on the standard
mixing-length prescription.

Many discoveries have been made that have
changed our understanding of our central G2 type star. Until a few
years ago, the Sun was thought to be lithium-poor by a factor of
10 compared to similar one-solar-mass solar-type disk stars
(Lambert \& Reddy 2004), which led to the suggestion
that the Sun is peculiar in its Li abundance and therefore
of dubious value for calibrating non-standard models of Li
depletion. However, current studies have shown that the Sun seems
to be typical when compared to solar twins instead of
merely solar-type stars, and that solar twins also have low
Li abundances (Mel\'endez  \& Ram\'irez 2007, Pasquini et al.
2008).

The history of internal mixing and Li depletion of each star
is probably dependent of the evolution of its rotational history
and the changes in the convective envelope over time. These
parameters are mass and metallicity dependent, and eventual
differences in them, from birth, may exist between very similar
stars and yet be too small to be measured observationally with
conventional techniques. Therefore, the detailed internal
modeling of solar twin stars, already established to be extremely
similar to the Sun in many astrophysical properties, is
potentially a powerful tool for boosting our comprehension of the
complex evolution of the Li abundance in low-mass stars.

We know that the Sun is not unique in being a planet host.
However, the Sun is still unique in the sense that no other
truly solar-like planetary system has been detected
to date, although some solar twins, which do not seem to
have either hot Jupiters or other giant planets in their
habitable zones (e.g., HIP 79672 (18 Sco), HIP 56948;
Mel\'endez \& Ram\'irez 2007), are excellent candidates for hosting
Earth-like planets, and therefore they should be analyzed in
detail both from the theoretical and observational point of view.
The quest to identify stellar analogues to the Sun or true solar twins
has continued for a long time (Cayrel de Strobel 1996,
Porto de Mello \& da Silva 1997). High resolution,
high signal-to-noise data analyzed by  Mel\'endez \&
Ram\'irez (2007) demonstrated that HIP 56948 is the most likely solar twin
known to date both in terms of stellar parameters and chemical
composition, including a low lithium abundance. The depletion of
Li in field solar-analog G dwarf stars appears to highlight our
limited understanding of the physics acting in the interiors of
stars. To explain the unexpected main-sequence Li
depletion of F and early G-type stars, a number of models and
extra-mixing processes have been proposed, including mass loss
(Swenson \& Faulkner 1992), diffusion (Michaud 1986; Chaboyer et
al. 1995), meridional circulation (Charbonnel \& Talon 1999 and
references therein), angular momentum loss, and rotationally
driven mixing (Schatzman \& Baglin 1991; Vauclair 1991;
Pinsonneault et al. 1992; Deliyannis \& Pinsonneault 1997;
Charbonnel \& do Nascimento 1998), gravity waves 
(Garc\'\i a L\'opez \& Spruit 1991; Montalban \&
Schatzmann 2000), tachocline (Brun et al. 1999; Piau et
al. 2003), and combinations of these (Charbonnel \& Talon 2005).
In this work, we present evolutionary solar-like models with
microscopic diffusion and rotation-induced mixing in the radiative
interior. This mixing due to the meridional circulation with a
feedback effect of the $\mu$-currents is described in Zahn (1992),
Vauclair \& Th\'eado (2003), and Th\'eado \& Vauclair (2003a). We
also introduce macroscopic motions caused by the tachocline (Richard
et al. 2004), where the shear turbulence mixes and homogenizes
with the convective zone the material transported by the
meridional circulation. The diffusion-circulation coupling
allowed both sides of the Li-dip to be reproduced (Th\'eado 
\& Vauclair 2003b) and the addition of a
tachocline provides very good agreement with helioseismic
observations in solar models (Richard et al. 2004). Since Li is
probably an indicator of the complex processes that occurred in the
past between the stellar external layers and the hotter interior,
its abundance can be used to identify true solar
twins, because solar twins are expected to share
not only the present parameters of the Sun but also the essential
aspects of its evolution.

The present working sample is described in Sect. 2, where we
redetermine the evolutionary status and individual masses of the
sample by using the HIPPARCOS parallaxes and by comparing
the observational Hertzsprung-Russell diagram with evolutionary
tracks. In Sect. 3, we describe the evolutionary models with
non-standard physics. In Sect. 4, we present the lithium abundance
main features and compare the observations with theoretical
predictions. Finally, our conclusions are outlined in Sect. 5.

\section{Solar twin sample}
\label{sec:Obs}

Our analysis is based on the three solar twins, HIP 55459, HIP
79672, and HIP 100963, analyzed by Takeda et al. (2007), and the
four solar twins, HIP 55459, HIP 79672, HIP 56948, and HIP
73815 studied by Mel\'endez \& Ram\'irez (2007). Two stars, HIP
55459 and HIP 79672, are present in both
samples. Thus, the sample consists of 5 solar twins.
Following the same procedure as do Nascimento et al. (2000), we
used the new HIPPARCOS trigonometric parallax measurements (van
Leeuwen 2007) to locate precisely the objects in the HR diagram.
Intrinsic absolute magnitudes M$_{\rm V}$ were derived from the
parallaxes and the m$_{\rm V}$ magnitudes were those given by HIPPARCOS.
We computed the stellar luminosity and the associated error from
the $\sigma$ error in the parallax. The uncertainties in luminosity
lower than $\pm0.1$ had an effect of $\pm0.03$ in the
determination of the masses.

 In Takeda et al. (2007), HIP 55459, HIP 79672, and HIP 100963 are taken from
the sample of 118 solar-analogue stars, observed by HIDES
(Izumiura 1999) at the coud\'e focus of the 188 cm reflector of
Okayama Astrophysical Observatory (OAO). These three stars are G 
main-sequence stars, which are considered to be  solar twins. Their Li 
abundance is however a factor of 3 to 4 higher than solar.
Mel\'endez \& Ram\'irez (2007) presented their data and results for  HIP 55459, HIP 79672, 
HIP 56948, and HIP 73815 observed with the 2dcoud\'e spectrograph
(Tull et al. 1995) on the 2.7-m Harlan J. Smith Telescope at
McDonald Observatory. HIP 56948 and HIP 73815 are considered by these 
authors to be the most similar stars to the Sun ever found, because  their Li abundances 
are as low as solar.
Atmospheric parameters of all of these stars are in excellent
agreement with those of the Sun, lower than 1.3\%  for the \teff~ and 
a few hundredths dex  for  $\log (L/L_{\odot})$ and [Fe/H].

Two observed trends provide a first order explanation
of the Li abundance in these solar twins. HIP
55459 and HIP 79672 are indeed found to be slightly more
massive than the Sun by $\approx$ 3-4 \% (Takeda et al. 2007;
Mel\'endez \& Ram\'irez 2007), which could  explain the 
overabundance of Li compared to solar lithium, the
external convective zone being shallower. The other trend concerns
the age of the stars since HIP 56948 and HIP 73815 are older than
the Sun and exhibit a stronger Li-depletion.  The sample of 
solar-twins is  summarized in Table \ref{tab:data_obs}.

%%%%%%%%%Figure 1 %%%%%%%%%%%%%%%%%%%
\begin{figure*}
\vspace{-0.1in}
\centerline{\psfig{figure=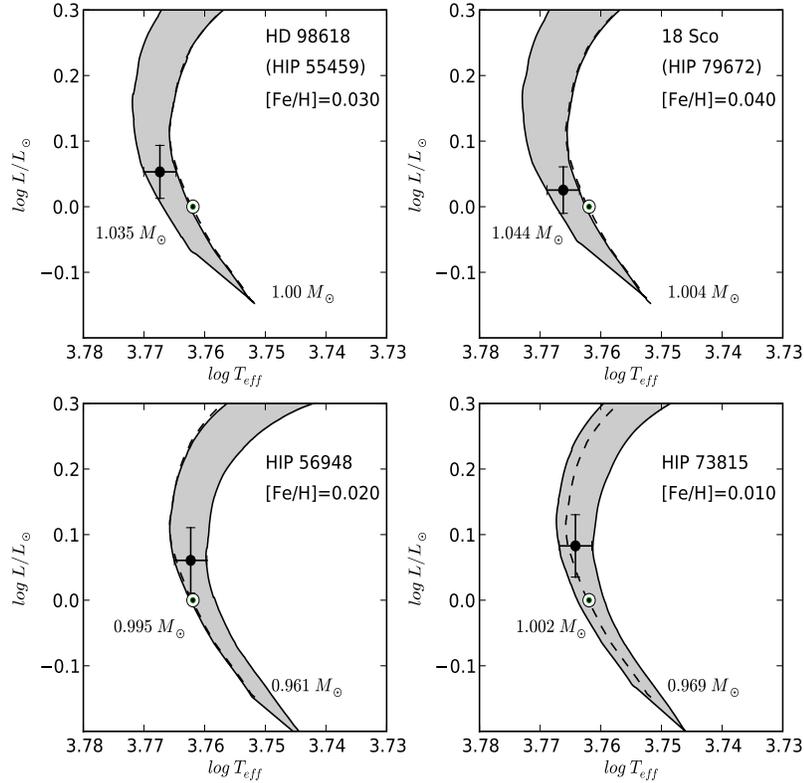,width=12cm,height=12cm}
\hskip 0.1in} \caption{Solar twins stars observed by M\'elendez \&
Ram\'irez (2007) in the Hertzsprung-Russell diagram. Luminosities
and related errors have been derived from the Hipparcos
parallaxes. The typical errors on \teff \ are indicated in
Table \ref{tab:data_obs}. The shaded zone represents the range of
masses in TGEC models (the maximum and the minimum masses are
indicated) limited by the 1-$\sigma$ observational error bars.}
\label{diaghr_Melendez}
\end{figure*}
%%%%%%%%% end Figure 1 %%%%%%%%%%%%%%%%%%%

\section{Stellar evolutionary models}

For the purposes of this study, stellar evolution calculations
were computed with the Toulouse-Geneva stellar evolution code TGEC
(Hui-Bon-Hoa  2008). Details of the physics of these models can be
found in Richard et al. (1996, 2004), do Nascimento et al. (2000),
and Hui-Bon-Hoa (2008). We describe the input standard
physics, non-standard processes, diffusion, and rotation-induced
mixing added in the models.

\subsection{Input physics}

We used the OPAL2001 equation of state by Rogers \& Nayfonov
(2002) and  the radiative opacities by Iglesias \& Rogers (1996),
completed with the low temperature atomic and molecular opacities
by Alexander \& Ferguson (1994). The nuclear reactions are from
the analytical formulae of the NACRE (Angulo et al. 1999)
compilation, taking into account the three \textit{pp} chains and
the CNO tricycle with the Bahcall \& Pinsonneault (1992)
screening routine. Convection is treated according to the
B\"ohm-Vitense (1958) formalism of the mixing length theory with a
mixing length parameter $\alpha = l/H_p =1.74$, where $l$ is
the mixing length and $H_p$ the pressure height scale. For the
atmosphere, we use a gray atmosphere following the Eddington
relation, which is a good approximation for main-sequence solar-type
stars (VandenBerg et al. 2008).

The abundance variations in the following chemical species are
computed individually  in the stellar evolution code: H, He, C, N,
O, Ne, and Mg. Both Li and Be are treated separately only as a fraction
of the initial abundance. The heavier elements are gathered in a
mean species Z. The initial composition follows the Grevesse \&
Noels (1993) mixture with an initial helium abundance
$\mathrm{Y_{ini}} = 0.268$. We chose to use the ``old" abundances
of Grevesse \& Noels (1993) instead of the ``new" mixture of
Asplund et al. (2005). This choice is motivated by the
disagreement between these new abundances  and the helioseismic
inversions for the sound-speed profile, the surface helium
abundance, and the convective zone depth. Furthermore, Caffau et al. 
(2009) revised the solar metallicity using 
3D hydrodynamical models to Z=0.0156 and Z/X=0.0213. These values 
are closer to those of Grevesse \& Noels (1993). \\
 
\noindent \textit{Diffusion and rotation-induced mixing} \\

The microscopic diffusion is computed with the atom test
approximation. All models include gravitational settling with
diffusion coefficients computed as in Paquette et al. (1986).
Radiative accelerations are not computed here, since we focus only on
solar-type stars where their effects are small when mixing is 
taken into account (Turcotte et al. 1998, Delahaye \& Pinsonneault 2005).
Rotation-induced mixing is computed as described in Th\'eado \&
Vauclair (2003a). This prescription is an extension of the approach of Zahn (1992)
and Maeder \& Zahn (1998), and introduces the feedback effect of
the $\mu$-currents in the meridional circulation, caused by  the
diffusion-induced molecular weight gradients. It introduces two
free parameters in the computations (C$\rm _h$ and $\alpha_{\rm
turb}$ : cf. Eq. (20) of Th\'eado \& Vauclair 2003a).
%%%%%%%%%Figure 2 %%%%%%%%%%%%%%%%%%%
\begin{figure}[h!]
\begin{center}
\vspace{-0.1in}
\hskip -0.3in
\includegraphics[angle=0,height=10cm,width=1.1\columnwidth]{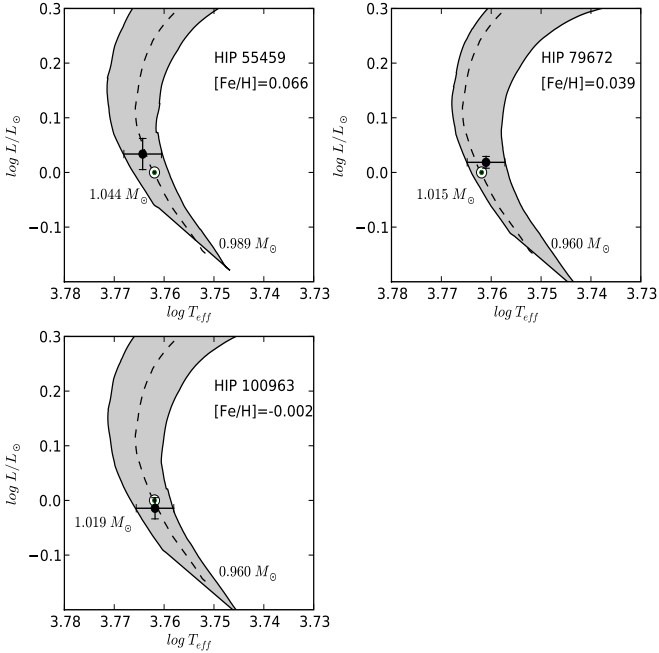}
\end{center}
\vspace{-0.2in}
\caption{ Solar twins observed by Takeda et al.
(2007), the caption is as in Fig. \ref{diaghr_Melendez}.} \label{diaghr_Takeda}
\end{figure}
%%%%%%%%% end Figure 2 %%%%%%%%%%%%%%%%%%%

The evolution in the rotation profile follows the Skumanich's law
(Skumanich 1972) with an initial surface rotation velocity on the
ZAMS equal to $V_i = 100$ km.s$^{-1}$. Other prescriptions were
tested by other authors to model the lithium destruction.
Charbonnel \& Talon (1999) and Palacios et al. (2003)
included angular momentum transport induced by mixing.
However, since a rotation-induced mixing alone cannot account for the
flat rotation profile inside the Sun, these authors later
introduced the possible effect of internal gravity waves triggered
at the bottom of the convective zone (e.g., Talon \& Charbonnel
2005), which allows us to  reproduce the hot side of the
Li-dip. Other authors suggested that the internal magnetic field is
more important than internal waves in transporting angular momentum
(Gough \& McIntyre 1998). In any case, when applied to the
solar case, all of these prescriptions are able to reproduce the
lithium depletion observed in the Hyades, and the results are
ultimately quite similar (Talon \& Charbonnel 1998, Theado
\& Vauclair 2003b).

We also include a shear layer below the convective zone, which is treated
as a tachocline (see Spiegel \& Zahn 1992). This layer is
parameterized with an effective diffusion coefficient that decreases
exponentially downwards (see Brun et al. 1998, Brun et al. 1999,
Richard et al. 2004):
\begin{displaymath}
D_{tacho} = D_{bcz} \exp \left( \ln 2 \frac{r - r_{bcz}}{\Delta} \right)
\end{displaymath}
where $D_{bcz}$ and $r_{bcz}$ are the value of
$D_{tacho}$ at the bottom of the convective zone and the radius at
this location respectively, and $\Delta$ is  the half width 
of the tachocline. Both $D_{bcz}$ and
$\Delta$ are free parameters and the absolute size
of the tachocline (i.e., $\Delta/R_*$ where R$_*$ is the radius of
the star) is supposed to be constant during the evolution.

\subsection{Models and calibration}

We calculated evolutionary models  of different masses from the
zero-age main sequence (ZAMS) to the end of the hydrogen
exhaustion in the core. Our evolutionary models were calibrated to
match the observed solar effective temperature and luminosity at
the solar age. The calibration method of the models is
based on the Richard et al.~(1996) prescription: for a 1.00
\msun~star, we adjusted the mixing-length parameter $\alpha$ and
the initial helium abundance Y$_{ini}$ to reproduce the observed solar
luminosity and radius at the solar age. The observed values that we
used are those of Richard et al.~(2004), i.e., L$_{\odot}$ = 3.8515 $\pm$ 0.0055
$\times$ 10$^{33}$ erg.s$^{-1}$, R$_{\odot}$ = 6.95749 $\pm$ 0.00241
$\times$ 10$^{10}$ cm, and age$_{\odot}$ = 4.57 $\pm$ 0.02 Gyrs. For 
best-fit solar model, we obtained L = 3.8499 $\times$ 10$^{33}$
erg.s$^{-1}$ and R = 6.95938 $\times$ 10$^{10}$ cm at an age = 4.576
Gyrs.

The free parameters of the rotation-induced mixing determine the
efficiency of the turbulent motions. They are adjusted to
produce a mixing that is both: 1) efficient and deep enough to smooth the
helium gradient below the outer convective zone; 2) weak and
shallow enough to avoid the destruction of Be. Following Grevesse
\& Sauval (1998), the Be abundance of the Sun is $\log$ N(Be) =
1.40 $\pm$ 0.09, and Balachandran \& Bell (1998) showed that, after
correcting for the continuous opacity in the ultraviolet region of
the spectrum, solar beryllium is not depleted at all with respect
to the meteoritic value. However, the source of ``extra UV" opacity has 
never been identified and it is justified by assuming that the OH lines 
in the UV and the IR are both formed in LTE. We obtained a slight Be 
destruction by a factor of 1.25 with respect to the meteoritic value, 
which is well within the error in the determination of the solar 
Be abundance.

The calibration of the tachocline allows us to reach the solar
lithium depletion ($\log$ N(Li) = 1.10 $\pm$ 0.10, e.g., Grevesse \&
Sauval 1998), and for our best-fit solar model we obtained $\log$
N(Li) = 1.13. We also checked  that the sound velocity profile of our best-fit 
model is consistent with that deduced from helioseismology
inversions by Basu et al. (1997). Our calibration is in an excellent 
agreement with helioseismology, more accurately
than 1\% for most of the star, except in the deep interior,
where it reaches 1.5\%. The input parameters for the other masses are 
the same as for the 1.00 \msun \ model.

\section{Results and discussion}
\label{sec:results}

Lithium is a key element because it is easily destroyed in stellar
interiors. Its abundance indicates the amount of internal mixing
in the stars and its destruction is strongly mass and age dependent.
Nowadays, it is well established on empirical arguments that a
non-standard mixing mechanism must be operating  to explain the low
Li abundances in solar-type stars.

%%%%%%%%%Figure 3 %%%%%%%%%%%%%%%%%%%
\begin{figure*}
\vspace{-0.1in}
\centerline{\psfig{figure=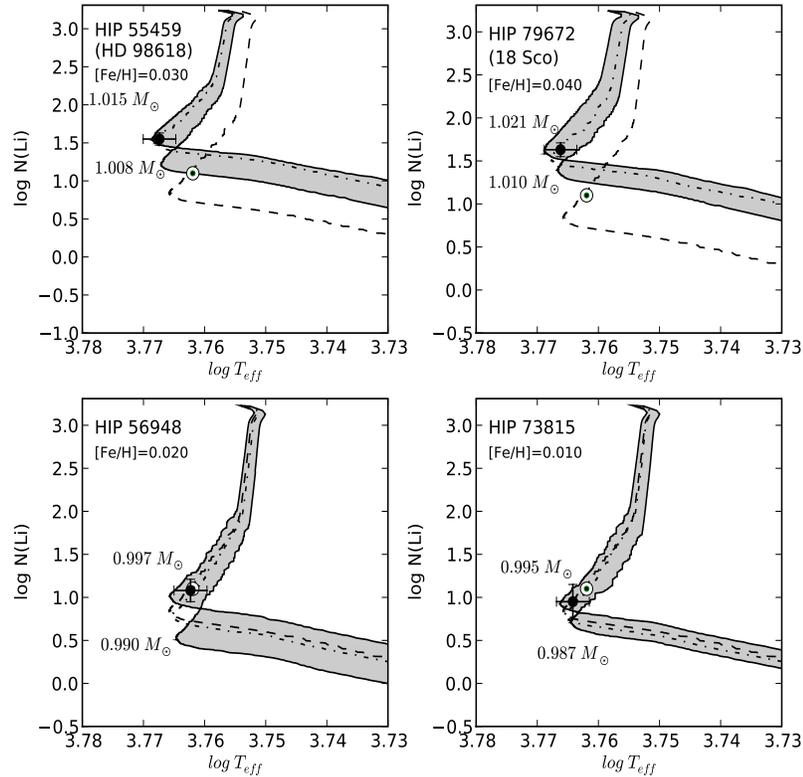,width=12cm,height=12cm}
\hskip 0.1in} \caption{Solar twins observed by M\'elendez
\& Ram\'irez (2007): lithium destruction along the
evolutionary tracks as a function of the effective temperature.
The shaded zone represents the range of masses of TGEC models (the
maximum and the minimum masses are indicated) limited by the
1-$\sigma$ observational error bars. A model (dot-dashed lines) of
1.013 \msun \ (HIP 55459), 1.015 \msun \ (HIP 79672), 0.994 \msun
\ (HIP 56948), and 0.992 \msun \ (HIP 73815) passes through the
observed point. The positions of the solar twin, the Sun,
and the lithium destruction of a solar model (dashed line) are
indicated.} \label{li-teff_Melendez}
\end{figure*}
%%%%%%%%%End Figure 3 %%%%%%%%%%%%%%%%%%%

For each star in our sample, we have first computed a grid of models as
described above with a mass step of 0.001 \msun~ and with
masses limited by the 1-$\sigma$ error bars in the HR diagram
(Figures \ref{diaghr_Melendez} and \ref{diaghr_Takeda}). The
metallicity of the models matches that observed for each
solar twin. This first comparison in the HR diagram between the
observations of solar twins and the models gives us an 
estimation of their mass and age, and of the precision of our
method. The masses inferred are within the range expected for 
the mass of a solar twin ($\pm$5 \% of the solar mass). The
precision  of the mass determination is directly linked to the
precision of the \teff \ estimations from the observations. In the
case of stars observed by M\'elendez \& Ram\'irez (2007), the
error associated with the effective temperature is $\pm$36 K, and
on average we obtain a spread in  masses $\mathrm{< \Delta M > \sim
0.036}$ \msun, within the limits of the \teff \ errors. For the 
stars studied by Takeda et al. (2007), the error associated with \teff
\ is $\pm$50 K and the average range of masses is $\mathrm{<
\Delta M > \sim 0.056}$ \msun. This first estimation is quite
satisfactory, but by using the observed Li abundance we are able to
reach even higher precision.

Figures \ref{li-teff_Melendez} and \ref{li-teff_Takeda} show the
lithium destruction of our models and the observed abundance of
each solar twin as a function of  effective temperature. The
error associated with $\log N{\rm (Li)}$ is about 0.1 dex. As in 
Figs. \ref{diaghr_Melendez} and \ref{diaghr_Takeda}, we plot a
zone corresponding to the grid of models with masses limited by the
1-$\sigma$ error bars. For each star, this infers a $\mathrm{\Delta
M}$ that is even smaller, $\mathrm{< \Delta M > \sim 0.008}$ \msun \ for
M\'elendez \& Ram\'irez (2007) stars, and $\mathrm{< \Delta M >
\sim 0.012}$ \msun \ for Takeda et al. (2007) stars. We also computed 
a model track that passes through the observed position of the star in the \teff~ -~ Li
 \ diagram. It provides the most probable modeling of the observed
star, and the values of our mass and age estimations.

 The uncertainties in the external parameters of our models
are difficult to evaluate and have different sources. The use of a more
sophisticated atmosphere model than the 1-D grey atmosphere
computations or a changed in the internal physics could modify the
derived effective temperature. However,  the present models and their
parameters were calibrated to account precisely for the
solar observed external parameters, i.e., the solar sound-velocity
profile and for the solar lithium depletion, with a solar model.
The uncertainties that we deduced for the mass and age estimations were
calculated for the values of mass and age from the two extreme
models in the diagrams Li-\teff. These uncertainties are lower 
limits since only two parameters are being sampled. In this 
context, the original result of our analysis is then that the
masses and  $\log N{\rm (Li)}$ values associated with a non-standard 
model at a given \teff \ provide us with a mass solution that is 
more precise than a mass determined based only on the HR diagram
position.

\subsection {Comparison with previous results}

%%%%%%%%%Figure 4 %%%%%%%%%%%%%%%%%%%
\begin{figure}[h!]
\begin{center}
\vspace{-0.1in}
\hskip -0.3in
\includegraphics[angle=0,height=10cm,width=1.1\columnwidth]{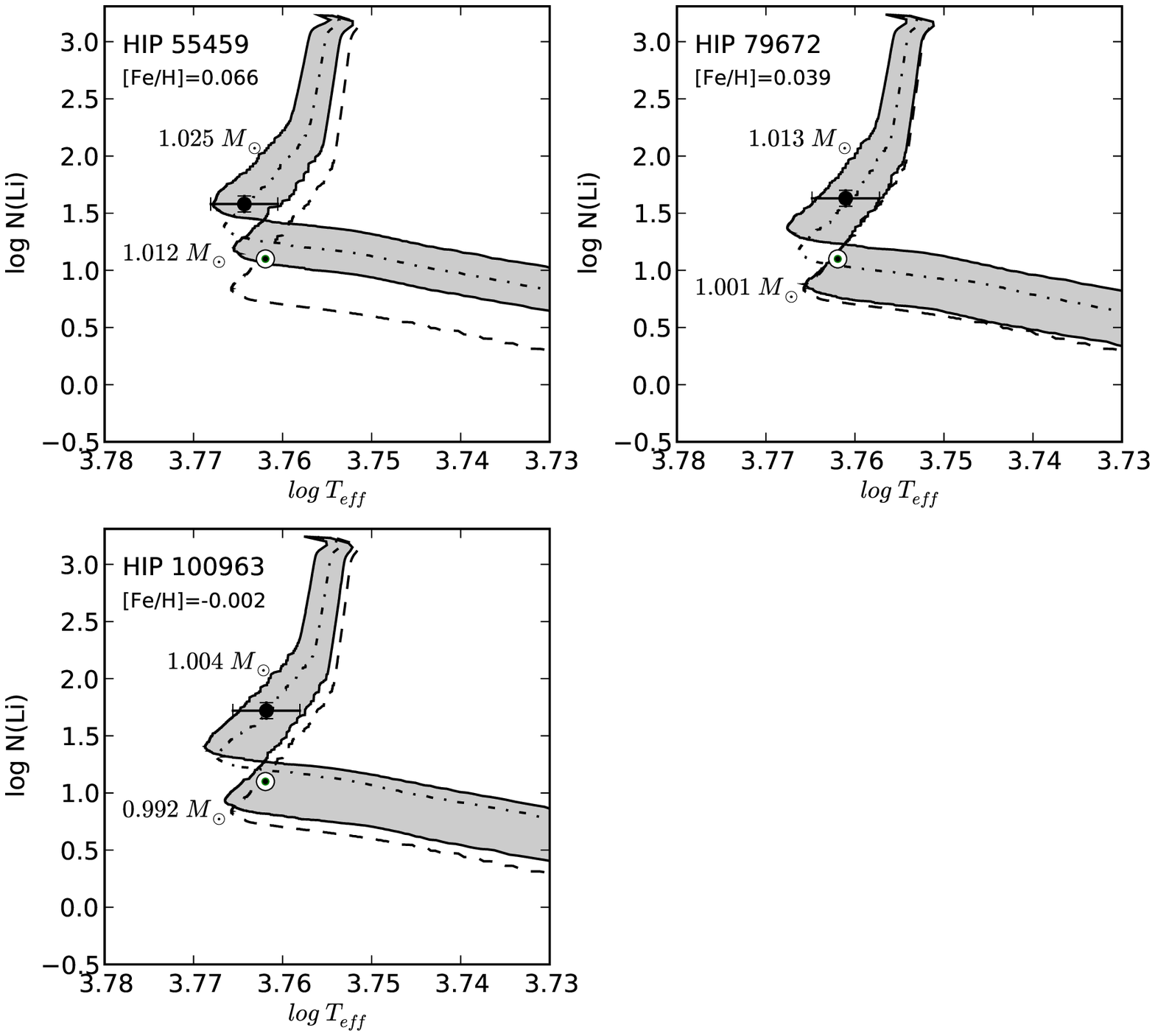}
\end{center}
\vspace{-0.2in}
\caption{Solar twins observed by Takeda et al.
(2007), the caption is as in Fig. \ref{li-teff_Melendez}.} \label{li-teff_Takeda}
\end{figure}

%%%%%%%%% end Figure 4 %%%%%%%%%%%%%%%%%%%

\noindent HIP 55459 was previously observed by Valenti \& Fisher
(2005) and Mel\'endez et al (2006). The values that we used in our analysis 
are those from Takeda et al. (2007) and  Mel\'endez \&
Ram\'irez (2007) (see Table \ref{tab:data_obs}). The Li abundance
appears to be enhanced compared to the Sun, although from 
observations and comparison with evolutionary models, the
star appears to be significantly more massive than the Sun with an
age between  $4.27 \pm 1.94$ Gyr and $5.77 \pm 0.81$ Gyr.
Our estimated mass, of $1.018 \pm 0.007$ \msun \ comparing with
Takeda et al. (2007) and $1.013 \pm 0.005$ \msun \ comparing with
Mel\'endez \& Ram\'irez (2007), is lower than all other
determinations (see Table \ref{tab:results}). However, in the
determination by comparison with the observations of Takeda et al.
(2007), we were unable to obtain an evolutionary track
consistent with the hot side of the error bar, as seen in
the left upper box of Fig. \ref{li-teff_Takeda}.

 HIP 79672 is the oldest star that is considered to be a
solar twin. It was observed by  Porto de Mello \& da Silva (1997), 
Valenti \& Fisher (2005), Luck \& Heiter (2005), and Mel\'endez et al.
(2006). In this work, we have used the values from both Takeda et al. (2007) 
and Mel\'endez \& Ram\'irez (2007) (see Table \ref{tab:data_obs}). 
The $\log N{\rm (Li)}$ of 18 Sco provides us with a picture of a star
with a mass between 0.7\%  and 1.5\% higher  than the Sun. 
The age is difficult to estimate because we found $2.89 \pm 1.09$ Gyr with
the observations of Takeda et al.(2007) and $5.03 \pm 1.29$
Gyr with Mel\'endez \& Ram\'irez (2007) (see
Table \ref{tab:results}). These  uncertainties in the present
estimations originate in the discrepancies between Mel\'endez \&
Ram\'irez (2007) and Takeda et al. (2007), especially for the
effective temperature estimation. In both cases, we can observe 
that all lithium determinations are in close agreement. Thus, the principal discrepancy
between mass and age estimated from the observations of Mel\'endez \&
Ram\'irez (2007) and Takeda et al. (2007) comes from the \teff~difference.

HIP 100963 was observed by  Masana et al.
(2006) and we used the observations of Takeda et al.
(2007) (see Table \ref{tab:data_obs}). The estimated age is around 
5.13 Gyr. Our estimation obtained by comparing with this most recent observation
implies a stellar mass  very close to the solar value (M = $0.998
\pm 0.006$ \msun) but younger (between 2.01 and 3.80 Gyr),
which is consistent with the observed lithium abundance (see
Table \ref{tab:results}).

\noindent HIP 56948 is currently the most likely solar twin.
Observations  of this star were completed  by Holmberg et al. (2007)
and Masana et al. (2006). The latest observations from 
Mel\'endez \& Ram\'irez (2007)  were used in this work
(see Table \ref{tab:data_obs}). We confirm that HIP 56948 appears to 
be an excellent solar-twin candidate of  mass M = $0.994 \pm 0.004$ \msun \ 
and  age = $4.71 \pm 1.39$ Gyr, which is even closer to the solar value than the age 
determined by Mel\'endez \& Ram\'irez (2007) ($5.8 \pm 1.0$ Gyr; see
Table \ref{tab:results}). If the mixing processes involved
in the interior are the same in both the Sun and HIP 56948, the fact
that these two stars have roughly the same Li content, and that
HIP 56948 is slightly less massive, suggests that the latter is
slightly younger.

\noindent HIP 73815 was analyzed by Robinson et
al. (2007).  We used the measurements of 
Mel\'endez \& Ram\'irez (2007) (see Table \ref{tab:data_obs}). The
abundance $\log N{\rm (Li)}$ is consistent with a solar twin of
$0.992 \pm 0.005$ \msun \ and an age of $5.76 \pm 1.13$ Gyr
(see Table \ref{tab:results}).

\begin{figure}[h!]
\begin{center}
\includegraphics[angle=0,height=11cm,width=\columnwidth]{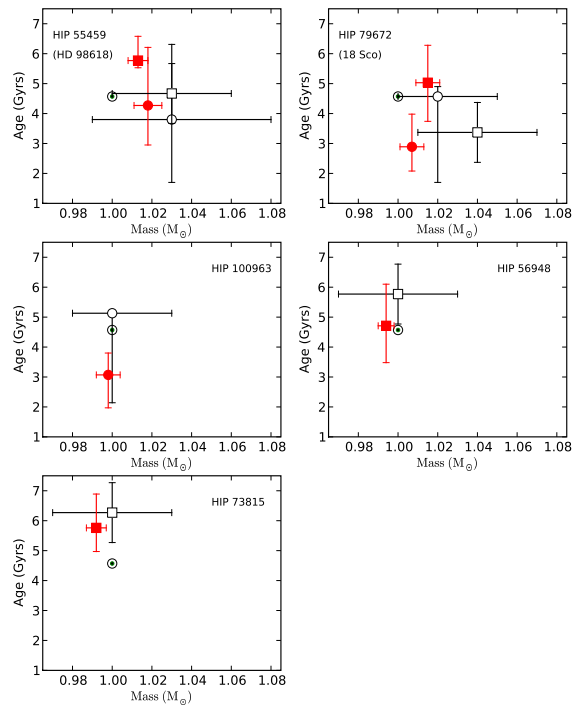}
\end{center}
\caption{Comparison between masses and ages determined by TGEC
models (filled symbols) and masses and ages estimated by the
observations (open symbols). Squares correspond to the solar twins
observed by M\'elendez \& Ram\'irez (2007) and circles to the solar twins
observed by Takeda et al. (2007). The errors bars are as described in the
text.} \label{fig:mass-age}
\end{figure}

Takeda \& Tajitsu (2009) proposed new
determinations of the external parameters of HIP 56948, HIP 79672, and HIP 100963 
(see Table \ref{tab:data_obs}). The new determination of \teff~ for
HIP 79672 is much closer to that of Mel\'endez \& Ram\'irez
(2007). A new estimation should reduce the discrepancies that we found
between the two determinations of both mass and age. The new
determination of \teff \ for HIP 100963 is close to the old value,  and
should not change  our estimation of mass and age significantly. Concerning HIP 56948,
these determinations are close to those  of Mel\'endez \&
Ram\'irez (2007), and a study that applies our method with these observations should confirm our
 results for this solar twin. Our results are summarized in Table \ref{tab:results} and 
 Fig. \ref{fig:mass-age}.\\

\section{Conclusions}

Although a fundamental parameter in studies of stellar
evolution, the mass of single stars cannot be derived directly from
observations. The mass is a crucial parameter in characterizing a solar twin.
We have analyzed the 5 most probable solar twins selected from
the literature and developed a method constrained by
the Li abundance observations to determine the mass and age
of the stars. Our mass determination is based on the
Li depletion in low-mass stars being  strongly mass dependent.
The Li depletion in solar-type stars is caused by  the
microscopic diffusion and the rotation-induced mixing. Lithium atoms
diffuse below the outer convective zone and if this zone is deep
enough, the meridional circulation transports them from the bottom
of the convective zone to the destruction layers. This process
provides  an explanation of the low Li abundance observed in
the Sun and some solar twins. We used accurate lithium abundance
observations compared with non-standard models to provide precise
information about the age and mass of these stars, which are more accurately
determined than by classical methods. For each solar twin, we
computed the non-standard model that gives the best-fit
solution to the observed lithium abundance. For HIP 56948, our analysis
infers  a mass of 0.994 $\pm$ 0.004 \msun \ and age of 4.71 $\pm$
1.39 Gyr. We have confirmed that HIP 56948 is the most likely solar twin at the
present time, its Li abundance being very close to the solar
value. HIP 73815 also has a low Li abundance and was proposed  by
Mel\'endez \& Ram\'irez (2007) as an excellent solar-twin candidate, although 
our analysis, constrained by its Li abundance, infers a star
slightly less massive and older than the Sun. HIP 55459 and 18 Sco
are found to be slightly more massive than the Sun, which is consistent
with their high Li abundance. HIP 100963 seems to be a young solar twin. 

In summary, we have shown that an accurate lithium abundance measurement and
 non-standard models provide more precise information
  about the age and mass of solar twins, more precisely than  determined 
 by classical methods.  This demonstrates that the physical
  properties of the stellar interior should be taken into consideration
  when attempting a more realistic characterization of a solar twin star.

\begin{table*}
\caption{Parameters of the observed solar twins.}
\label{tab:data_obs}
\centering
\begin{minipage}[t]{\textwidth}
\centering
\begin{tabular}{cccccccl}
\hline
Name & $T_{eff}$ & $(L/L_{\odot})$ & [Fe/H] & $\log N$(Li) & Age & Mass & source\footnote{The sources of the observations are: (a) Porto de Mello \& da Silva (1997), (b) Luck \& Heiter (2005), (c)  Valenti \& Fisher (2005), (d) Masana et al. (2006), (e) Mel\'endez et al (2006), (f) Holmberg et al. (2007), (g)* Mel\'endez \& Ram\'irez (2007), (h) Robinson et al. (2007), (i)* Takeda et al. (2007), (j) Takeda \& Tajitsu (2009). The observations used in this work are represented by *.} \\
 & (K) & & & & (Gyrs) & (\msun) & \\
\hline
HIP 55459 & $5812 \pm 50$ &  $1.081 \pm 0.069$ & 0.066 & $1.58 \pm 0.07$ & $3.80^{+2.51}_{-2.10}$ & $1.03^{+0.05}_{-0.04}$ & (i)* \\
\\
(HD 98618) & $5837 \pm 36$ & $1.13 \pm 0.11$ & 0.030 & $1.55 \pm 0.08$ & $4.7 \pm 1.0$ & $1.03 \pm 0.03$ & (g)* \\
\\
%%%%%%%%%%
 & $5812 \pm 44$ & (...) & $0.03 \pm 0.03$ & (...) & $4.9 \pm 2.9$ & $1.040 \pm 0.150$ &  (c) \\
 \\
 & $5843 \pm 30$ & (...) & $0.05 \pm 0.03$ & $1.57 \pm 0.09$ & $4.3 \pm 0.9$ & $1.02 \pm 0.03$ & (e) \\
\hline
HIP 79672 & $5762 \pm 50$ &  $1.043 \pm 0.027$ & 0.039 & $1.63 \pm 0.07$ & $4.57^{+0.33}_{-2.87}$ & $1.02^{+0.03}_{-0.02}$ & (i)* \\
\\
(18 Sco) & $5853 \pm 36$ & $1.06 \pm 0.09$ & 0.040 & $1.63 \pm 0.08$ & $3.4 \pm 1.0$ & $1.04 \pm 0.03$ & (g)* \\
\\
 & $5789 \pm 30$ & (...) & $0.05 \pm 0.06$ & (...) & (...) & (...) & (a) \\
 \\
 & 5823 & (...) & 0.03 & (...) & (...) & (...) & (b) \\
 \\
 & $5791 \pm 44$ & (...) & $0.03 \pm 0.03$ & (...) & $4.7 \pm 2.7$ & $0.980 \pm 0.130$ & (c) \\
 \\
 & $5817 \pm 30$ & (...) & $0.02 \pm 0.03$ & $1.63 \pm 0.09$ & $4.0 \pm 0.4$ & $1.02 \pm 0.03$ & (e) \\
 \\
 & 5815 & (...) & 0.047 & 1.60 & (...) & (...) & (j) \\
\hline
HIP 100963 & $5779 \pm 50$ & $0.968 \pm 0.043$ & -0.002 & $1.72 \pm 0.07$ & $5.13^{+0.00}_{-2.99}$\footnote{Takeda et al. (2007) give no estimation for the upper error of the age of HIP 100963.} & $1.00^{+0.03}_{-0.02}$ & (i)* \\
\\
 & 5794 & (...) & (...) & (...) & (...) & (...) & (d) \\
 \\
 & 5803 & (...) & -0.004 & 1.68 & (...) & (...) & (j) \\
\hline
HIP 56948 & $5785 \pm 36$ &  $1.15 \pm 0.14$ & 0.020 & $1.08 \pm 0.13$ & $5.8 \pm 1.0$ & $1.00 \pm 0.03$ & (g)* \\
\\
 & 5785 & (...) & (...) & (...) & (...) & (...) & (d) \\
 \\
 & 5701 & (...) & -0.15 & (...) & 9.6 & $0.98 \pm 0.05$ & (f) \\
 \\
 & 5791 & (...) & 0.020 & 1.13 & (...) & (...) & (j) \\
\hline
HIP 73815 & $5810 \pm 36$ &  $1.21 \pm 0.14$ & 0.010 & $0.95 \pm 0.20$ & $6.3 \pm 1.0$ & $1.00 \pm 0.03$ & (g)* \\
\\
 & 5759 & (...) & -0.05 & (...) & (...) & (...) & (h) \\
\hline
\end{tabular}
\end{minipage}
\end{table*}

\begin{table*}
\caption{Mass and age determinations from TGEC models compared to
observations of (a) Takeda et al. (2007) and (b) Mel\'endez \&
Ram\'irez (2007)} \label{tab:results} \centering
\begin{tabular}{ccccccc}
\hline
Name & \multicolumn{3}{c}{Mass [\msun]} & \multicolumn{3}{c}{Age [Gyr]} \\
 & (a) & (b) & (TGEC) & (a) & (b) & (TGEC) \\
\hline
HIP 55459 & $1.03^{+0.05}_{-0.04}$ & & $1.018 \pm 0.007$ & $3.80^{+2.51}_{-2.10}$ & & $4.27^{+1.94}_{-1.32}$ \\
\\
(HD 98618) & & $1.03 \pm 0.03$ & $1.013 \pm 0.005$ & & $4.7 \pm 1.0$ & $5.77^{+0.81}_{0.24}$ \\
\\
\hline
HIP 79672 & $1.02^{+0.03}_{-0.02}$ & & $1.007 \pm 0.006$ & $4.57^{+0.33}_{-2.87}$ & & $2.89^{+1.09}_{-0.81}$ \\
\\
(18 Sco) & & $1.04 \pm 0.03$ & $1.015 \pm 0.06$ & & $3.4 \pm 1.0$ & $5.03^{+1.25}_{-1.29}$\\
\\
\hline
HIP 100963 & $1.00^{+0.03}_{-0.02}$ & & $0.998 \pm 0.006$ & $5.13^{+0.00}_{-2.99}$ & & $3.07^{+0.73}_{-1.06}$ \\
\\
\hline
HIP 56948 & & $1.00 \pm 0.03$ & $0.994 \pm 0.004$ & & $5.8 \pm 1.0$ & $4.71^{+1.39}_{-1.23}$ \\
\\
\hline
HIP 73815 & & $1.00 \pm 0.03$ & $0.992 \pm 0.005$ & & $6.3 \pm 1.0$ & $5.76^{+1.13}_{-0.79}$ \\
\\
\hline
\end{tabular}
\end{table*}

\begin{acknowledgements}
This research has made use of SIMBAD and VIZIER databases,
operated at CDS (Strasbourg, France). JRM thanks support from the
FCT (Ciencia 2007).
Research activities of the Stellar
Board at the Federal University of Rio Grande do Norte are supported
by continuous grants from CNPq and FAPERN Brazilian Agencies.
GFPM acknowledges financial support by CNPq and FAPERJ.
The authors acknowledge support from the FCT/CAPES cooperation
agreement n$^{o}$237/09.
\end{acknowledgements}

\bibliographystyle{aa}

{}

\end{document}